\documentclass[12pt]{emulateapj}

\def\lesssim{\mathrel{\hbox{\rlap{\hbox{\lower4pt\hbox{$\sim$}}}\hbox{$<$}}}}
\def\gtrsim{\mathrel{\hbox{\rlap{\hbox{\lower4pt\hbox{$\sim$}}}\hbox{$>$}}}}

\def\ks{\kappa_{\rm T}}

\def\rhoo{\rho_{\rm 0}}

\def\tg{T_{\rm g}}
\def\teff{T_{\rm eff}}
\def\am{a_{\rm mag}}
\def\amo{a_{\rm 0}}
\def\qo{Q_{\rm 0}}
\def\rhoo{\rho_{\rm 0}}
\def\fd{f_{\rm d}}
\def\so{\Sigma_{\rm 0}}
\def\B{\begin{equation}}
\def\E{\end{equation}}
\def\Fo{F_{\rm 0}}
\def\B{\begin{equation}}
\def\E{\end{equation}}
\def\O{\Omega}
\def\fd{f_{\rm d}}
\def\ft{f_{\rm t}}
\def\kb{k_{\rm B}}
\def\me{m_{\rm e}}
\def\te{\tau_{\rm e}}

\usepackage{graphicx}

\slugcomment{}

\begin{document}

\title{Effects of Local Dissipation Profiles on Magnetized Accretion Disk Spectra}

\shortauthors{Tao \& Blaes}

\author{Ted Tao}
\affil{Department of Physics, St. Mary's College of Maryland, St. Mary's City MD 20686}

\author{Omer Blaes}
\affil{Department of Physics, University of California,
Santa Barbara CA 93106}

\begin{abstract}

We present spectral calculations of non-LTE accretion disk models appropriate for high luminosity stellar mass black hole X-ray binary systems. We first use a dissipation profile based on scaling the results of shearing box simulations of \cite{hir08} to a range of annuli parameters. We simultaneously scale the effective temperature, orbital frequency and surface density with luminosity and radius according to the standard $\alpha$-model \citep{ss73}. This naturally brings increased dissipation to the disk surface layers (around the photospheres) at small radii and high luminosities. We find that the local spectrum transitions directly from a modified black body to a saturated Compton scattering spectrum as we increase the effective temperature and orbital frequency while decreasing midplane surface density. Next, we construct annuli models based on the parameters of a $L/L_{\rm Edd}=0.8$ disk orbiting a $6.62$ solar mass black hole using two modified dissipation profiles that explicitly put more dissipation per unit mass near the disk surface. The new dissipation profiles are qualitatively similar to the one found by \cite{hir08}, but produce strong near power-law spectral tails. Our models also include physically motivated magnetic acceleration support based once again on scaling the \cite{hir08} results. We present three full-disk spectra each based on one of the dissipation prescriptions. Our most aggressive dissipation profile results in a disk spectrum that is in approximate quantitative agreement with certain observations of the steep power law (SPL) spectral states from some black hole X-ray binaries. 

\end{abstract}


\section{Introduction}

Understanding the internal structure of radiating accretion disks remains a major outstanding problem in astrophysics and is the key for connecting theoretical models to the observed spectra from systems such as black hole X-ray binaries and active galactic nuclei (AGN). In a series of papers, \cite{h98}, \cite{h00}, and \cite{h01} adapted the one-dimensional time-independent stellar atmosphere code TLUSTY for accretion disk applications. These authors constructed non-LTE accretion disk annuli models that self-consistently solved the radiative transfer and vertical structure equations. The results were appropriate for supermassive black holes and eventually took into account relativistic effects, Comptonization and metal opacities. Despite their complexities, these models still have significant physical limitations. In particular, they employed an ad-hoc prescription where dissipation per unit mass as a function of height was assumed to be constant. 

Recent vertically stratified shearing box simulations have led to significant progress in understanding the internal structure of accretion flows. These calculations evolve the time dependent three-dimensional radiation magnetohydrodynamic equations in a box embedded inside the disk and account for the vertical tidal gravity due to the central black hole. Such simulations have been completed for a wide range of box-integrated radiation to gas pressure ratios of interest \citep{tur04, hir06, kro07, bla07, hir08}. The resulting spatial dissipation profiles generally peak at slightly less than a pressure scale height away from the disk midplane \citep{bla11}. While the underlying physics is still unclear, these dissipation distributions presumably capture the effects of vertically stratified magnetorotational \citep{bh91, bh98} turbulence. Moreover, these studies showed that magnetic pressure support against gravity may be significant or even dominant (in the outer layers) compared to radiation pressure support. A limitation of the shearing box simulations is that they rely on flux limited diffusion for treating photons and do not solve the full radiative transfer equation. 

Studies that self-consistently integrate full multi-frequency radiative transfer into the time-dependent three-dimensional magnetohydrodynamic simulations do not yet exist, though significant progress toward that goal is being made \citep{jia12,dav12}. Many authors have instead attempted to incorporate physics results from the shearing box simulations into TLUSTY calculations. For example, \cite{dav05} used the time and horizontal averaged dissipation profile as a function of height from the \cite{tur04} simulation. Compared to a model based on the ad hoc constant dissipation per unit mass prescription, the resulting annulus vertical structure is very different but the emergent spectrum is not. \cite{bla06} included a horizontal and time averaged magnetic acceleration profile derived from the \cite{hir06} simulation. They found that magnetic pressure support produced a disk atmosphere with a larger vertical density scale height and a harder spectrum. More recently, \cite{dav09} utilized both the magnetic acceleration and dissipation profiles from \cite{hir08} in a TLUSTY calculation. The resulting disk spectrum is once again harder than the case without magnetic support but similar to models with the ad hoc dissipation profile. These authors also performed a separate Monte Carlo radiative transfer calculation that propagated photons through a three dimensional snap shot of the simulation domain. They discovered that the complicated density inhomogeneities in the photospheric regions actually soften the spectrum and largely negate the hardening that resulted from including the significant magnetic pressure support against gravity. Note that all of the above studies relied on the $\alpha$ model to generate radial disk profiles in order to produce full disk spectra.

The stratified shearing box simulations by \cite{hir08} found no evidence of the radiation pressure dominated thermal instabilities predicted by the standard $\alpha-$model \citep{ss76}.  However, more recent simulations by \citet{jia13} find evidence of thermal runaways.  Moreover, these disks may still be vulnerable to viscous (inflow) instabilities \citep{le74, hir09}, so whether such flows are ultimately stable is still open to debate. Nevertheless, if we assume that radiation pressure dominated disks can exist in nature, they provide a viable mechanism for powering the high luminosity steep power law (SPL, also known as very high) and soft (also known as high) states observed in accreting systems.
\cite{hir09} find that the thermal equilibrium relation between surface density and time-averaged (over many thermal times) vertically integrated stress actually agrees with the one predicted by the standard $\alpha$ prescription. Furthermore, these authors found that the time averaged and vertically integrated total thermal pressure (radiation plus gas) does correlate closely with the vertically integrated stress, once again in agreement with the $\alpha$ model.

These results suggest that the $\alpha$ model predictions for the relationships between basic annuli parameters such as surface density $\Sigma$, accretion rate $\dot{M}$ (which is proportional to luminosity), effective temperature $T_{\rm eff}$ and distance $r/r_{\rm g}$ (where $r_{\rm g}$ is the gravitational radius $=GM/c^2$) from the black hole may actually hold, at least in a time averaged sense. In particular, the surface density of a radiation pressure dominated disk annulus will decrease with increasing accretion rate and decreasing distance from the black hole. Additionally, the disk half thickness, which is proportional to its pressure scale height, will increase with higher accretion rate. Moreover, the shearing box simulations collectively revealed that the locations of the (effective and scattering) photospheres and the spatial distribution of turbulent dissipation are both tied to the scale height. While both the dissipation profile and photospheres move vertically outward at higher accretion rates, the simulations also suggest that the photospheres actually move closer to the dissipation peaks as the surface density drops and effective temperature climbs. This means that the fraction of accretion power dissipated near the spectral forming photospheric region increases as we raise the annulus luminosity and move towards the black hole. At high accretion rates, there may be enough dissipation in the upper layers of those annuli close to the black hole to power a corona that can Compton up-scatter a significant number of photons to very high energies. This might be a plausible mechanism for generating the high energy steep power law spectra. Previously, \cite{dav05} tried a dissipation profile that is discontinuous at the scattering photosphere, which resulted in dissipating $50$ percent of the flux above the scattering photosphere as well as a power-law like annulus spectrum that qualitatively resembled the SPL.  

In this work, we explore this possibility and extend these previous efforts by comparing the effects of three different dissipation profiles on the disk vertical structure and emergent spectrum. The first dissipation profile is a fit to that produced by the shearing box simulations of \cite{hir08}. The two others explicitly put more dissipation per unit mass in the low column density regions. We also include simulation motivated magnetic acceleration profiles in our calculations. We first study the effects of increased dissipation (of turbulent energy into heating the gas) in the disk upper layers as luminosity increases and annulus radius decreases. Next, we compare the effects of the three dissipation prescriptions on the spectrum of a $L/L_{\rm Edd}=0.8$ disk. We organize the paper as follows. In section 2, we summarize our numerical method for solving the vertical accretion disk structure plus radiative transfer equations. Section 3 presents some theoretical estimates regarding the disk structure to help us better understand our results. We then move on to presenting our calculated spectra and disk structures in section 4. In section 5, we outline the key physics underlying our numerical results. Finally, we discuss the astrophysical implications of our findings in section 6 and conclude in section 7. 

\section{Method}
We use the computer program TLUSTY \citep{hl95} to compute non-LTE time-independent one dimensional disk annuli vertical structure models and emergent spectra. We incorporate dissipation profiles based on simulations of the turbulence
generated by the magnetorotational instability \citep{bh91} that serves as the physical source of stress. The code self-consistently solves the disk structure equations along with angle-dependent radiative transfer but treats electron (both Thomson and Compton) scattering with an angle-averaged Kompaneets source function \citep{h01}. We ignore the effects of potentially important physics such as outflow and irradiation. Since the mean free path for electron-ion collisions is generally very short compared to the disk scale height, we assume that the electrons and protons are thermally locked, so that ionized gas can be locally characterized by a single temperature $T_{\rm g}$. We assume that energy transport is purely vertical and is only by diffusive radiative flux so that the dissipation $Q$ is related to the frequency-integrated flux $F$ via
\B
Q(z)=\frac{dF}{dz},
\label{qdfdz}
\E
where $z$ is the height measured vertically upward so that $z=0$ is the disk mid-plane. \cite{bla11} discovered that this is unlikely to be the case in nature as the latest shearing box simulations revealed that radiation pressure work and magnetic buoyancy driven radiation advection also contribute significantly to vertical energy transport in radiation pressure dominated disks. In particular, radiation advection can be as important as the radiative diffusion flux near the mid-plane. However, in the \cite{bla11} simulations, both radiation pressure work and radiation advection flux become small compared to radiative diffusion flux at higher altitudes, so that in the end it is radiative diffusion that carries the dissipated energy into the photospheric region where the spectrum forms, regardless of how energy transport occurs deep inside the disk. We therefore suspect that ignoring radiation advection will not significantly change our spectral results.

\begin{figure}
\includegraphics[width=9cm]{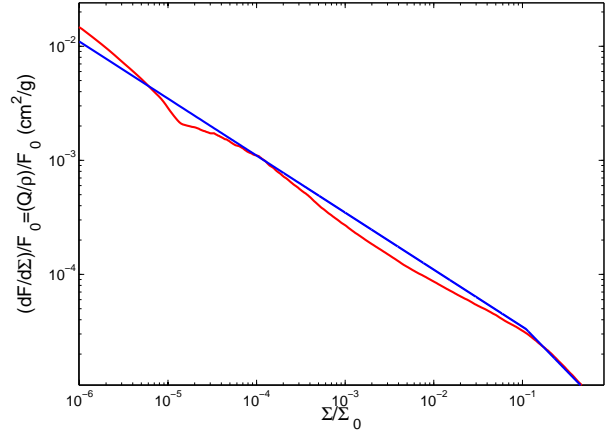}
\caption{Dissipation per unit mass normalized to total annulus flux as a function of fractional surface density. The figure compares the broken-power law prescription (blue) given by equation (\ref{dis1}) to the actual horizontally averaged dissipation profile from simulation $1112a$ (red) of \cite{hir08}. Instead of plotting equation (\ref{dis1}) (which gives $Q\Sigma/\rho$) directly, note that here we actually used $Q/\rho$ to highlight the dissipation per unit mass in the outer layer.}
\label{fig:q1112a}
\end{figure}

To  generate input for TLUSTY, we express the dissipation profile as a function of fractional surface density. In the first case, we fit the horizontal and time averaged dissipation profile from \cite{hir08} with a broken power law of the form
\B
\frac{Q\Sigma}{\rho}=-\Sigma\frac{dF}{d\Sigma}=F_0\left\{\begin{array}{ccc}A\left(\frac{\Sigma}{\so}\right)^{0.5}&,& \Sigma/\so<0.11\\B\left(\frac{\Sigma}{\so}\right)^{0.2}&,& \Sigma/\so>0.11\end{array}\right..
\label{dis1}
\E
as shown in Figure \ref{fig:q1112a}. Here $\Fo$ and $\so$ denote the total outgoing flux and mid-plane surface density, respectively. Here $A\approx 0.65$ and $B\approx 0.33$ are normalization constants computed automatically in the code to ensure that the profile is continuous at $\Sigma/\Sigma_0=0.11$ and that the total flux is $F_0$ so that
\B
F_0=\int_0^{\Sigma_0}\frac{dF}{d\Sigma} \ d\Sigma.
\E
The surface density $\Sigma$ at height $z$ is given by
\B
\Sigma(z)=\int_{z}^{\infty}\rho(z')dz',  
\E 
where $\rho$ is the depth dependent density. Note that $\rho$ and $\Sigma$ both go to zero at infinity since we defined $z=0$ to be the disk mid-plane. \cite{dav09} used a similar profile with a power law index of $0.5$ throughout the disk. Here we chose to fit the simulation in slightly more detail, resulting in the break at $\Sigma/\Sigma_0=0.11$. 

\begin{figure}
\includegraphics[width=9cm]{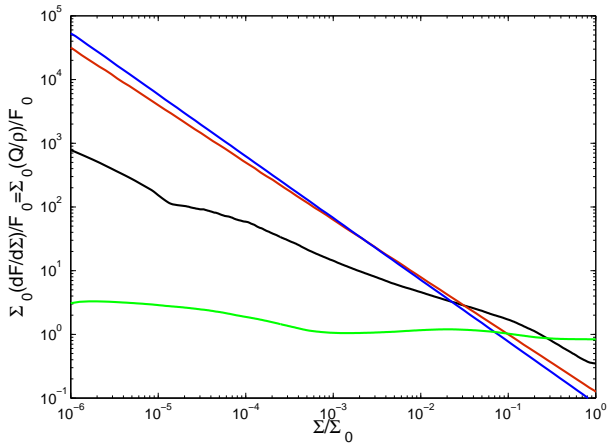}
\caption{Dissipation per unit mass normalized to total annulus flux times mid-plane surface density as a function of fractional surface density. The black curve represents the dissipation profile from simulation $1112a$ of \cite{hir08}, which we model with equation (\ref{dis1}). The blue and red curves represent the single-power law profiles of equation (\ref{dis2}) with $\zeta=0.03$ and $\zeta=0.1$, respectively. In both cases, we used an annulus at $R/R_{\rm g}=40.3$ of the $L/L_{\rm Edd}=0.8$ disk. Note that the aggressive $\zeta=0.03$ profile puts even more power into the disk upper layers than the moderate $\zeta=0.1$ prescription. For comparison, the green curve represents the dissipation profile based on the $\alpha$-prescription, where we took $Q=-\tau_{\rm r\phi}d\Omega/d{{\rm ln} \ r}$, and used a vertically averaged $\alpha\approx 0.02$ according to results from \cite{hir09}. As in Figure \ref{fig:q1112a}, here we plotted $Q/\rho$ instead of $Q\Sigma/\rho$.}
\label{fig:disp}
\end{figure}

Our other dissipation profiles are given by
\B
\frac{Q\Sigma}{\rho}=-\Sigma\frac{dF}{d\Sigma}=F_0\zeta\left(\frac{\Sigma}{\so}\right)^{\zeta},
\label{dis2}
\E
where the power-law index $\zeta$ is equal to $0.1$ or $0.03$. From here on, we shall refer to the $0.1$ case as the moderate profile and the $0.03$ case as the aggressive profile because the latter dissipates an even higher fraction of the accretion power in the disk upper layers around the photospheres. Note that in both cases, we have artificially increased the dissipation per unit mass $Q/\rho=dF/d\Sigma$ at low surface densities compared to equation (\ref{dis1}).  These dissipation profiles are all shown in Figure~\ref{fig:disp}.  As it turns out, our calculation shows that these simpler but somewhat artificial prescriptions still result in dissipation profiles $Q(z)$ that qualitatively agree with \cite{hir08} when viewed as a function of height, as shown in Figure~\ref{fig:z0.9q}. In practice, we manually imposed a cut-off for $\Sigma<10^{-6}\so$ to these single power law dissipation profiles in order to ensure energy conservation (in other words, to ensure that the total flux produced by the models is equal to $\sigma \teff^4$) within the calculation domain. 

Figure~\ref{fig:disp} also illustrates the dissipation profile that would result
from a vertically {\it local} alpha model, in which dissipation locally tracked
thermal pressure as a function of height.  This is inconsistent with
simulations, and puts far less dissipation above the photosphere than our other
profiles.

Our key assumption is that the dissipation per unit mass always has the same functional dependence on fractional surface density $\Sigma/\so$ regardless of annuli parameters. Given $\Sigma_0$, $\Omega$ and $\teff$, the total flux $\Fo$ ensures the correct normalization for the dissipation profile while the power law dependence on fractional surface density encodes the depth dependence. We will elaborate further on the implications of this assumption in the next section.

To incorporate magnetic support against gravity, we assume the magnetic acceleration profile of an annulus is tied to the pressure scale height $H$ and simply
expands vertically as we increase $H$ by increasing $\teff$ and decreasing $\so$. In other words, the magnetic acceleration profile has the form
\begin{equation}
\am(x)=\amo f(x),
\label{eqnam}
\end{equation}
where $x=z/H$ is the vertical position in an annulus normalized to the scale height, and $\amo$ is the mid-plane acceleration. To get $f(x)$, which represents the height dependence of $\am$, we smoothed time and horizontally averaged magnetic acceleration data from simulation $1112a$ of \cite{hir09}. We assumed that only the mid-plane magnetic acceleration $\amo$ changes with annuli parameters in our models. This assumption is consistent with magnetic acceleration profiles from other recent simulations \citep{hir08, hir09}. To obtain $\amo$, we assume that the magnetic acceleration simply scales with the radiative acceleration so that $\amo$ is proportional to $\kappa \Fo/c$, where $\Fo=\sigma\teff^4$ is the total (frequency integrated) emergent flux from the annulus surface and $\kappa$ is the total (frequency integrated) opacity. In other words, the relative importance of magnetic and radiative accelerations remain unchanged in our disk models as we sweep through the parameter space. We should note that the shearing box accretion disks simulations mentioned earlier have radiation support against gravity becoming more important compared to the magnetic contributions as the radiation to gas pressure ratio of a disk segment is increased. This means that our prescription will likely produce harder spectra \citep{bla06} than what the highest radiation to gas pressure ratio simulation would predict. However, \cite{dav09} found that this hardening is largely compensated by softening due to density inhomogeneities.

We first study the evolution of  annuli spectra with mid-plane surface density $\so$, orbital frequency $\Omega$ and effective temperature $T_{\rm eff}$. Our focus  is on low surface density (down to $801 \rm g/cm^2$ ) and high effective temperature (up to $1.81\times 10^7 \rm K$) annuli that might represent conditions within the inner parts of high luminosity disks around stellar mass black holes. 

Next, we construct the full spectrum for a disk around a $6.62$ solar mass Kerr black hole with spin $a/M=0.5$ using both of the above dissipation profiles. We divide the disk into $52$ individual annuli extending down to $5R_{\rm g}$ and out to $60R_{\rm g}$, where $R_{\rm g}=GM/c^2$ is the gravitational radius. Following previous authors, we compute $\alpha$ model-based $\so$ and $\teff$
radial profiles using the fully relativistic one-zone disk radial structure equations by \cite{nt73}, including the \cite{rh95} corrections. We have also employed the standard no-torque inner boundary conditions at the inner most stable circular orbit (ISCO) of the disk (but see, e.g., \citealt{kro05}). We then use a relativistic transfer function code \citep{agol97} to generate the full disk spectrum as seen by observers at infinity. Finally, we ignore the potentially important effects of radial advection and assume that energy transport is purely vertical.
A proper study of these effects requires two-dimensional slim disk models that couple the equations of radial and vertical structure  \citep{sa11} instead of simply computing the radial profile from height-averaged quantities. For our work, we checked that vertical radiative diffusion time is significantly less than the infall time for all annuli that we modeled. 

\section{Theoretical Expectations}

In order to better understand the physics behind our numerical results, we first explore the likely consequences of changing $\so$, $\O$ and $\teff$ by estimating the variation of photosphere locations, both scattering and effective, with these parameters. We assume a geometrically thin disk in the radiation pressure dominated limit to obtain the basic annuli parameters. To make the calculation easier, we re-parametrize the problem in terms of variations in annulus radial distance $r$ from the black hole and in disk luminosity $L$. For simplicity, we ignore general relativistic corrections for our simple estimates, although all numerical calculations elsewhere in this paper include them. Our annuli parameters are related to $L$ and $r$ via
\B
\teff=\left(\frac{3GML_{\rm E}}{8\pi\sigma\eta c^2}\right)^{1/4}\left(\frac{L}{L_{\rm E}}\right)^{1/4}\left(\frac{1}{r}\right)^{3/4}
\label{teffr}
\E
and
\B
\O=\left(\frac{GM}{r^3}\right)^{1/2}.
\label{or}
\E
Here $\eta$, $r_{\rm g}$, $L_{\rm E}$ and $\ks$ denote the accretion efficiency, gravitational radius, Eddington luminosity and Thomson opacity, respectively. Equation (\ref{teffr}) arises out of the fact that $F\propto L/r^3$ if the annulus is far away from the black hole and is thus in the non-relativistic regime. Finally, for a radiation pressure dominated annulus, the pressure scale height is on the order of (and is proportional to) the disk half-thickness, and is therefore given by
\B
\frac{H}{r}=\frac{3\kappa}{2\ks\eta}\frac{L}{L_{\rm E}}\frac{r_{\rm g}}{r}.
\label{hr}
\E
The equations above do not make any assumptions about the local turbulent stresses. However, to obtain scalings for the surface mass density, we must assume a stress prescription, and we adopt the $\alpha$-model \citep{ss73}, which in turn implies
\B
\so=\frac{8}{3\alpha}\frac{\eta}{L/L_{\rm E}}\left(\frac{\kappa}{\kappa_{\rm T}}\right)^{-2}\left(\frac{r}{r_{\rm g}}\right)\frac{\mu_{\rm e}}{\sigma_{\rm T}}.
\label{sr}
\E

To proceed, consider a semi-infinite plane parallel atmosphere with only free-free and electron scattering opacities. Our numerical models elsewhere in the paper include bound-free opacities and Compton scattering. We define the lower boundary to be at $z=0$. The effective and scattering photospheres occur at depths $z_e$ and $z_s$ such that
\begin{equation}
\tau_{\rm e} (z_{\rm e})\approx\int_{z_{\rm e}}^\infty{\rho(z)\sqrt{\kappa_{\rm T}\kappa_{\rm \rm ff}(z)}dz}=1
\label{taue}
\end{equation}
and
\begin{equation}
\tau_{\rm s} (z_{\rm s})=\int_{z_{\rm s}}^\infty{\rho(z)\kappa_{\rm T}dz}=1,
\label{taus}
\end{equation}
where
\begin{equation}
\kappa_{\rm ff}(z)=\frac{A\rho(z)}{T_{\rm g}(z)^{7/2}}
\end{equation}
is the free-free opacity with $A$ being a constant. Here $\rho$ and $T_{\rm g}$ are the gas density and temperature, respectively. Also, $\tau_{\rm e}$ and $\tau_{\rm s}$ are the frequency averaged effective and scattering optical depths of the media, respectively.

Next, for a disk annulus with pressure scale height $H$, we assume that the density and gas temperature, like magnetic acceleration, are also only functions of $x=z/H$, so that
\begin{equation}
\rho(x)=\rhoo f_{\rm d}(x)
\label{eqnd}
\end{equation}
and
\begin{equation}
\tg (x)=T_{\rm 0} f_{\rm t}(x),
\label{eqnt}
\end{equation}
where $\rhoo$ and $T_{\rm 0}$ denote the density and temperature at the disk mid-plane, respectively. The functions $f_{\rm d}(x)$ and $f_{\rm t}(x)$ are thus invariant under annuli parameter changes and encode the height dependence of density and temperature.

Recall our prescription that the dissipation per unit mass has the same functional dependence on fractional surface density regardless of annuli parameters. This assumption together with equation (\ref{eqnd}) means that we can also write the local dissipation as a function of height in the form of
\B
Q(z)=\qo q(x),
\label{eqnq}
\E
where the function $q(x)$ is once again independent of annuli parameters. The normalization $\qo$ depends on annuli parameters through the fact that $Q=dF/dz$ and $F\propto L/r^3$. In particular, this means that $\qo$ only changes with $r$ but not $L$, at least far away from the black hole where we can ignore relativistic effects. The more important implication of equation (\ref{eqnq}) is that the position of the dissipation profile peaks observed in the \cite{hir08} simulations remains at the same $x_{\rm p}=z_{\rm p}/H$ regardless of annuli parameters. In other words, we have assumed that the dissipation profile is tied to the disk scale height $H$ and simply stretches out as we increase $H$, which is in reasonable agreement with the simulation results in \cite{bla11}.

We begin with a disk annulus with luminosity $L_{\rm 1}$ at a distance $r_{\rm 1}$ away from the black hole and scale to a disk at some new $L$ and $r$. Equations (\ref{or})-(\ref{sr}) imply
\begin{equation}
\rhoo=\rho_{\rm 1}\left(\frac{L_{\rm 1}}{L}\right)^2\left(\frac{r}{r_{\rm 1}}\right)^{3/2},
\end{equation}
\begin{equation}
T_{\rm 0}=T_{\rm 1}\left(\frac{r}{r_{\rm 1}}\right)^{-3/8}
\end{equation}
and
\begin{equation}
H=H_{\rm 1}\frac{L}{L_{\rm 1}},
\end{equation}
where $\rho_{\rm 1}$, $T_{\rm 1}$ and $H_{\rm 1}$ are the mid-plane density, temperature and scale height of the original disk annulus, respectively. The temperature scaling arises from the total radiative flux $F(r)\propto L/r^3$ and the assumption that radiative diffusion is the dominant mechanism of vertical heat transport. Inserting these scaling relations into equations (\ref{taue}) and (\ref{taus}), and changing variable from $z$ to $x$, yields
\begin{equation}
\tau_{\rm e}(x_{\rm e})\approx \frac{H_{\rm 1}\rho_{\rm 1}^{3/2}\sqrt{\ks A}}{T_{\rm 1}^{7/4}}\left(\frac{L_{\rm 1}}{L}\right)^2\left(\frac{r}{r_{\rm 1}}\right)^{\frac{51}{32}}\int_{x_{\rm e}}^\infty\sqrt{\frac{\fd(x)^3}{\ft(x)^{7/2}}}dx=1
\label{tauex}
\end{equation}        
and
\begin{equation}
\tau_{\rm s}(x_{\rm e})=H_{\rm 1}\ks\frac{L_{\rm 1}}{L}\left(\frac{r}{r_{\rm 1}}\right)^{3/2}\int_{x_{\rm e}}^\infty \fd(x)dx=1.
\label{tausx}
\end{equation}
In deriving the above we assumed once again that $f_{\rm d}(x)$ and $f_{\rm t}(x)$ are invariant under changing $L$ and $r$. Equations (\ref{tauex}) and (\ref{tausx}) imply that both $x_{\rm e}$ and $x_{\rm s}$ must decrease as we raise the disk luminosity and move towards to the black hole. In contrast, the dissipation profile peak stays at the same $x=x_{\rm p}$ independent of $L$ and $r$ as discussed above. Hence the spectral forming photospheric region will move inwards, at least in scale height units, towards segments of higher dissipation with increasing $L$ and decreasing $r$ (or equivalently, increasing $\O$, $\teff$ and decreasing $\so$).

This result suggests that the fraction of accretion power dissipated in the spectral forming region (around the photospheres) will increase as we increase $\O$, $\teff$ and decrease $\so$. The higher dissipation rate may drive the electron temperature to rise sharply over the photon temperature immediately beyond the point where the gas and radiation thermally decouple. Such a vertical temperature structure may result in a hot corona that Compton up scatters photons from the colder, optically thick disk underneath. This disc-corona geometry has been invoked to explain the steep power law (SPL, also known as very high) state spectra commonly observed in black hole X-ray binary systems \citep{kd04}. As we shall see in section 5 below, for this scenario to work, the electron temperature gradient at the disk-corona interface needs to be steep enough to produce a downward sloping power law spectrum.

\section{Results}

\subsection{Spectra and Vertical Structure}

\begin{figure}
\includegraphics[width=9cm]{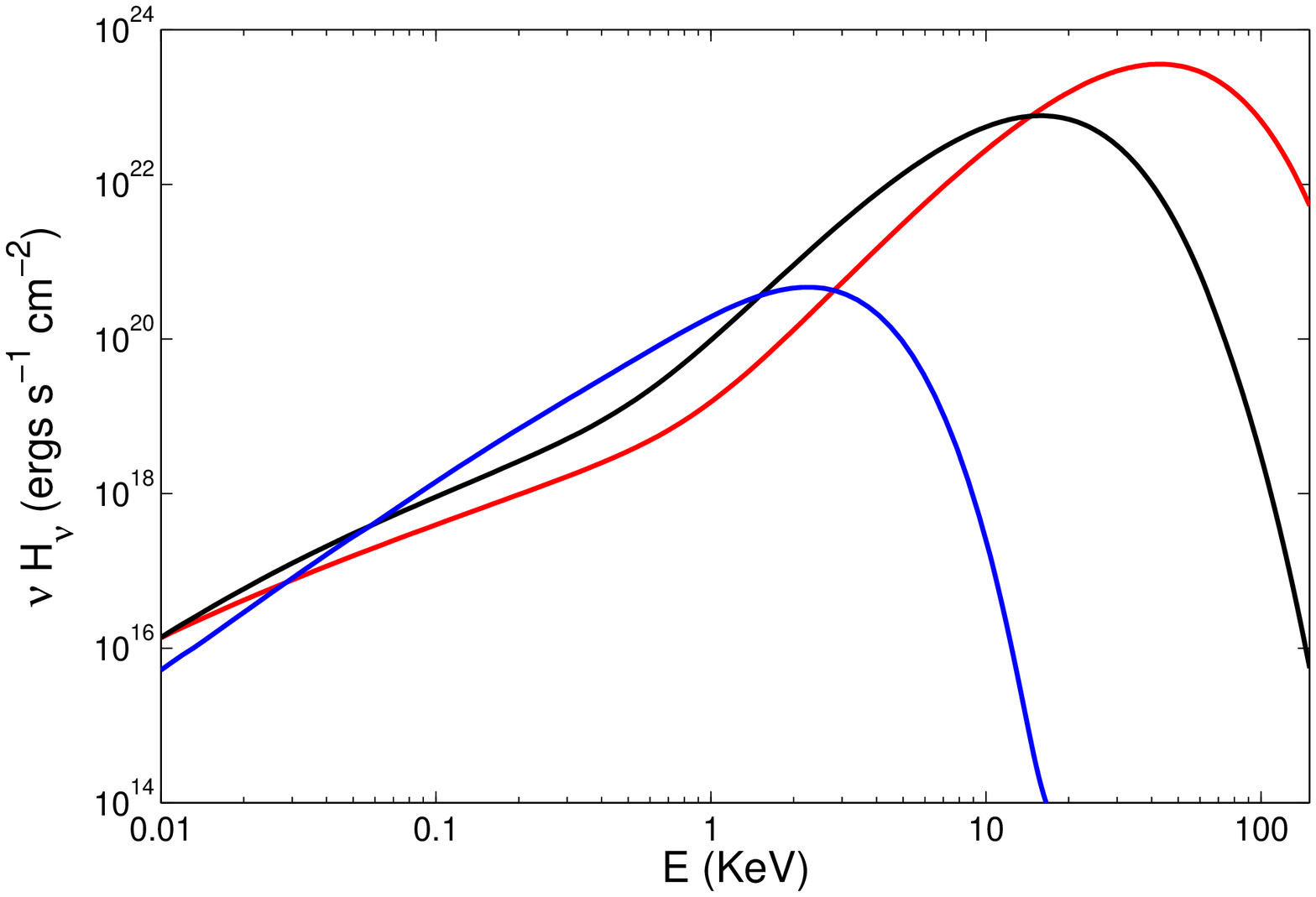}
\includegraphics[width=9cm]{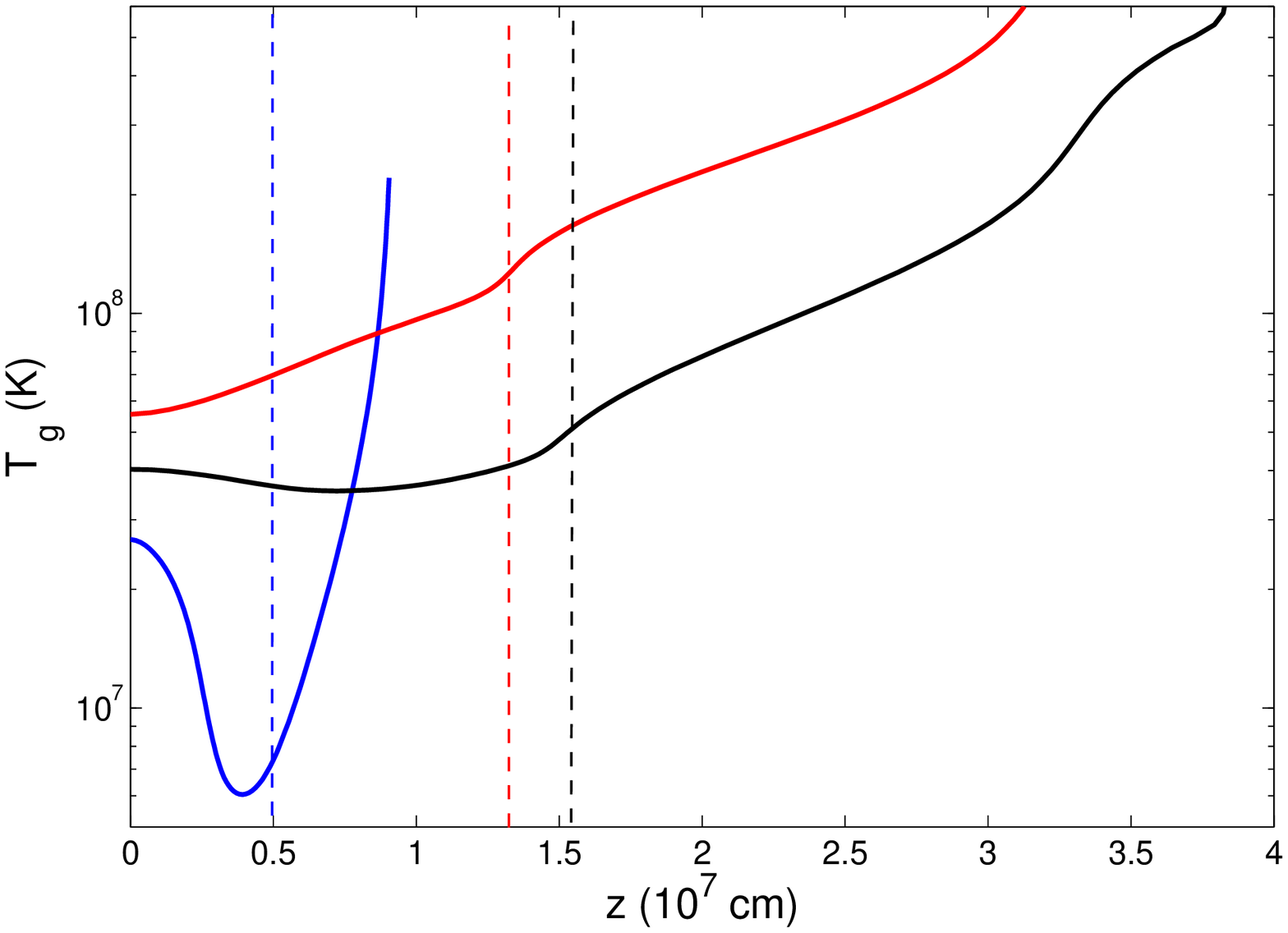}
\caption{Gas temperature (bottom panel) and emergent spectrum (top panel) from annuli models selected to illustrate the evolution of spectra with disk parameters. All models here use the broken power law dissipation profile from equation (\ref{dis1}). The blue curves correspond to parameters from the \cite{hir08} simulation with $\teff=3.46\times 10^6 \rm \ K$, $\O^2=3.61\times 10^4 \rm \ s^{-2}$ and $\so=5.31\times 10^4 \rm \ g/cm^2$. The black curve has $\teff=1.23\times 10^7 \rm \ K$, $\O^2=9.75\times 10^5\rm \ s^{-2}$ and $\so=1722 \rm \ g/cm^2$. Finally, the red curve has $\teff=1.81\times 10^7 \rm \ K$, $\O^2=4.51\times 10^6 \rm \ s^{-2}$ and $\so=801 \rm \ g/cm^2$. The vertical dashed lines on the lower panel indicate the locations of the scattering photospheres for each model.}
\label{fig:annuli}
\end{figure}

Figure \ref{fig:annuli} shows the vertical structures and resulting spectra for a series of three annuli models using the broken power law dissipation profile from equation (\ref{dis1}). Recall that this dissipation prescription is based on the results from the simulations of \cite{hir08}. To better understand these models, we calculated a frequency-dependent Compton $y$-parameter by integrating outwards from the effective photosphere at each frequency. For models that show prominent saturated Compton scattering ($T_{\rm eff}=1.81\times 10^7 \ \rm K$ and $T_{\rm eff}=1.23\times 10^7 \ \rm K$), the frequency-dependent Compton $y$-parameter is much less than $1$ at low frequencies but rises sharply at higher frequencies as scattering begins to dominate over absorption opacity. The model with the original \cite{hir08} simulation parameters, however, does not display significant Comptonisation and its $y$-parameter is much less than unity at all frequencies. The maximum $y$-parameter increases (from about $0.01$ to $10$ and then $20$) as we increase $\teff$, $\O$ and decrease $\so$, in agreement with the evolution of the spectral shapes in this plot. Note that the lowest surface density ($801\rm \ g/cm^2$) annulus is actually marginally effectively thin, while all other models presented here are effectively thick.

\begin{figure}
\includegraphics[width=9cm]{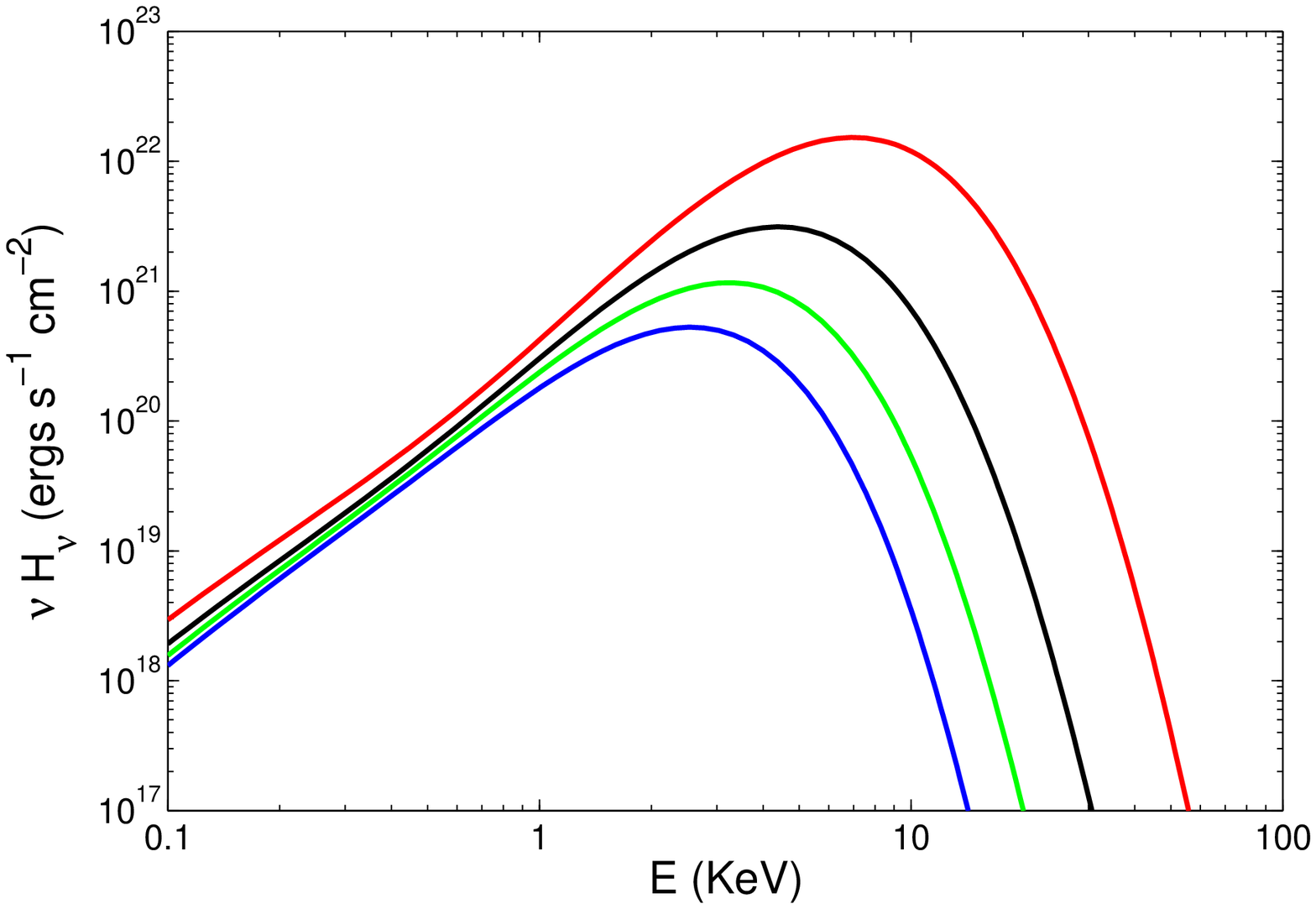}
\includegraphics[width=9cm]{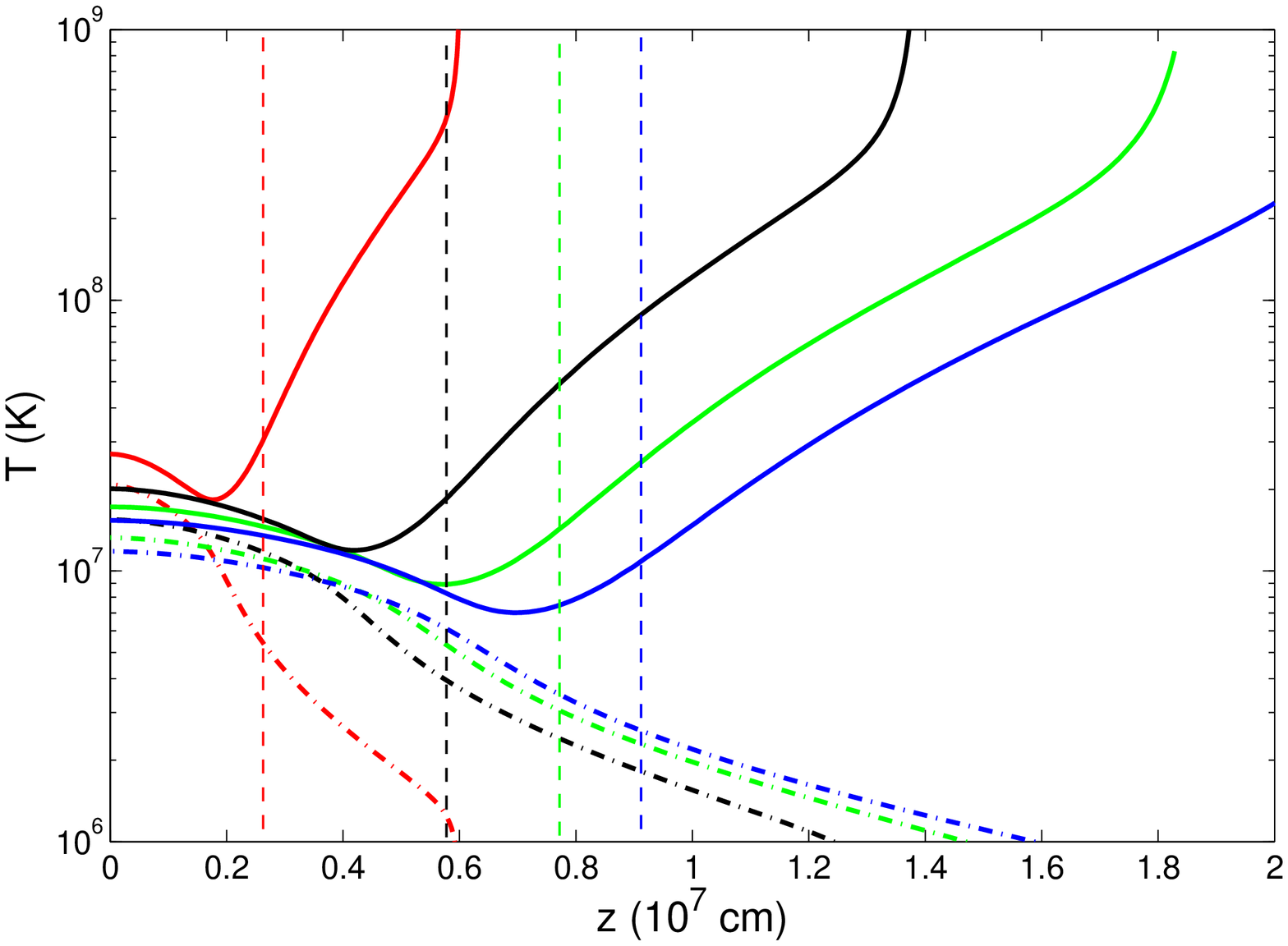}
\caption{Annuli spectrum and vertical structure due to the broken power law dissipation profile from equation (\ref{dis1}). The top panel shows emergent spectrum from annulus at $r/r_{\rm g}=40$ (blue), $30$ (green), $20.2$ (black) and $9.67$ (red). The bottom panel plots the gas (solid) and radiation (dot-dash) temperatures of the models, along with the locations of the corresponding scattering photospheres (dash).}
\label{fig:z0.8}
\end{figure}

In Figure \ref{fig:z0.8}, we show results representative of the series of annuli models that we use later to construct a disk integrated spectrum utilizing the same broken power law dissipation profile. The disk has $L/L_{\rm Edd}=0.8$ and orbits a black hole with spin parameter $a/M=0.5$. Note that none of the models in Figures \ref{fig:annuli} and \ref{fig:z0.8} display a significant high energy spectral tail despite covering a wide range of $T_{\rm eff}$, $\Omega$ and $\Sigma_0$, indicating that simply changing these basic disk parameters does not produce spectra consistent with the SPL state.



\begin{figure}
\includegraphics[width=9cm]{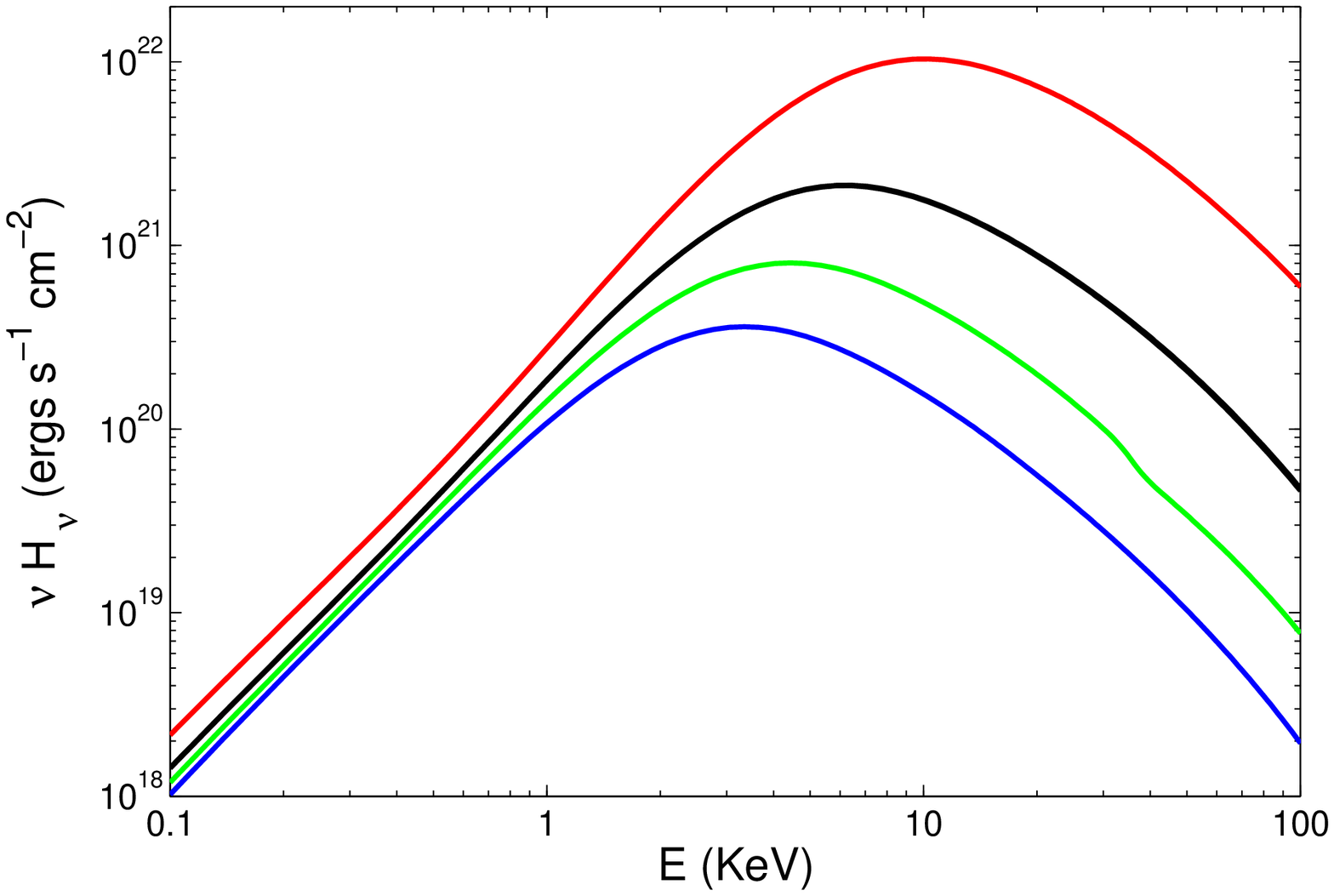}
\includegraphics[width=9cm]{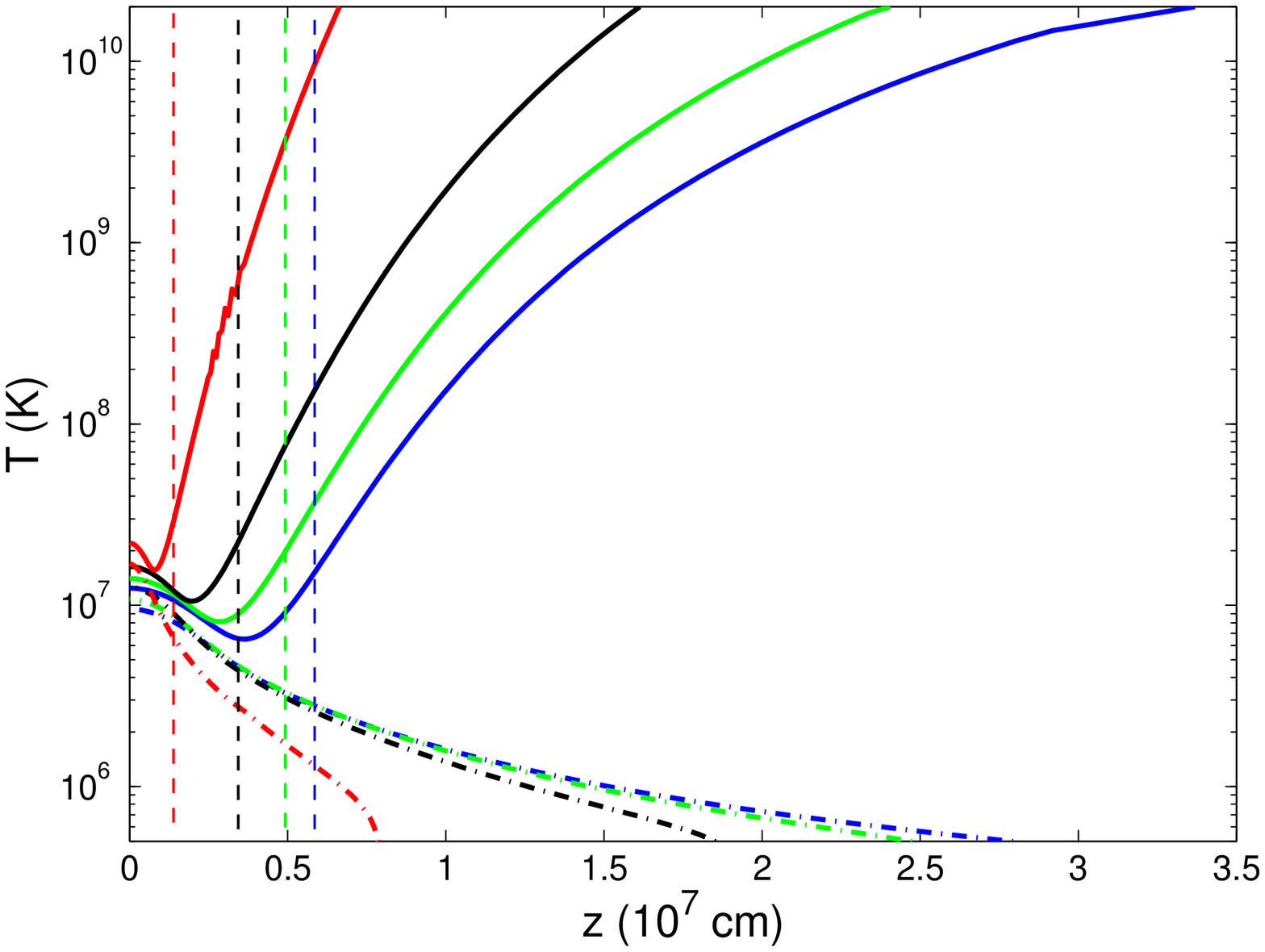}
\caption{Annuli spectrum and vertical structure due to the $\zeta=0.1$ single power law dissipation profile from equation (\ref{dis2}). The top panel shows emergent spectrum from annulus at $r/r_{\rm g}=40$ (blue), $30$ (green), $20.2$ (black) and $9.67$ (red). The bottom panel plots the gas (solid) and radiation (dot-dash) temperatures of the models, along with the locations of the corresponding scattering photospheres (dash).}
\label{fig:z0.9}
\end{figure}

\begin{figure}
\includegraphics[width=9cm]{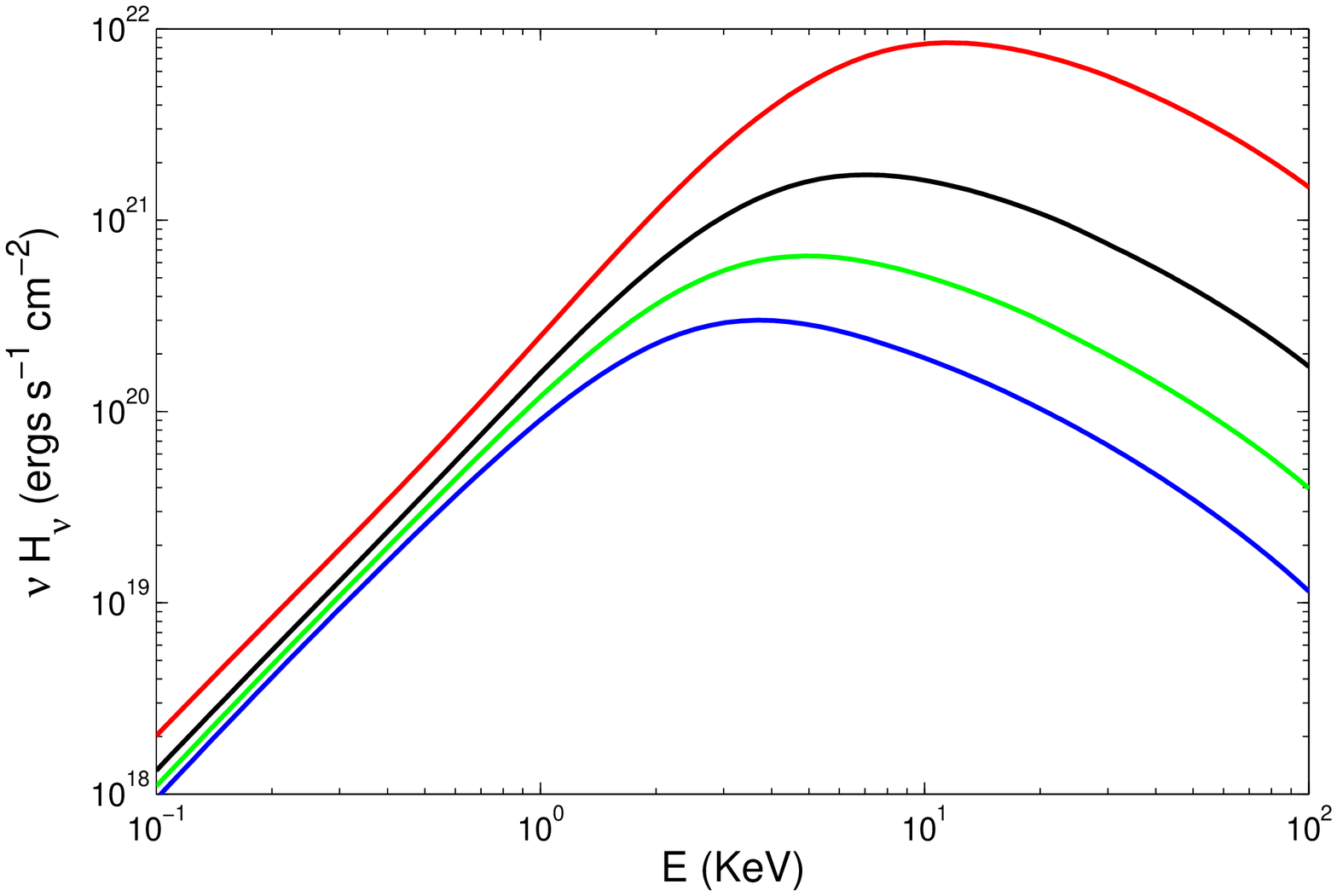}
\includegraphics[width=9cm]{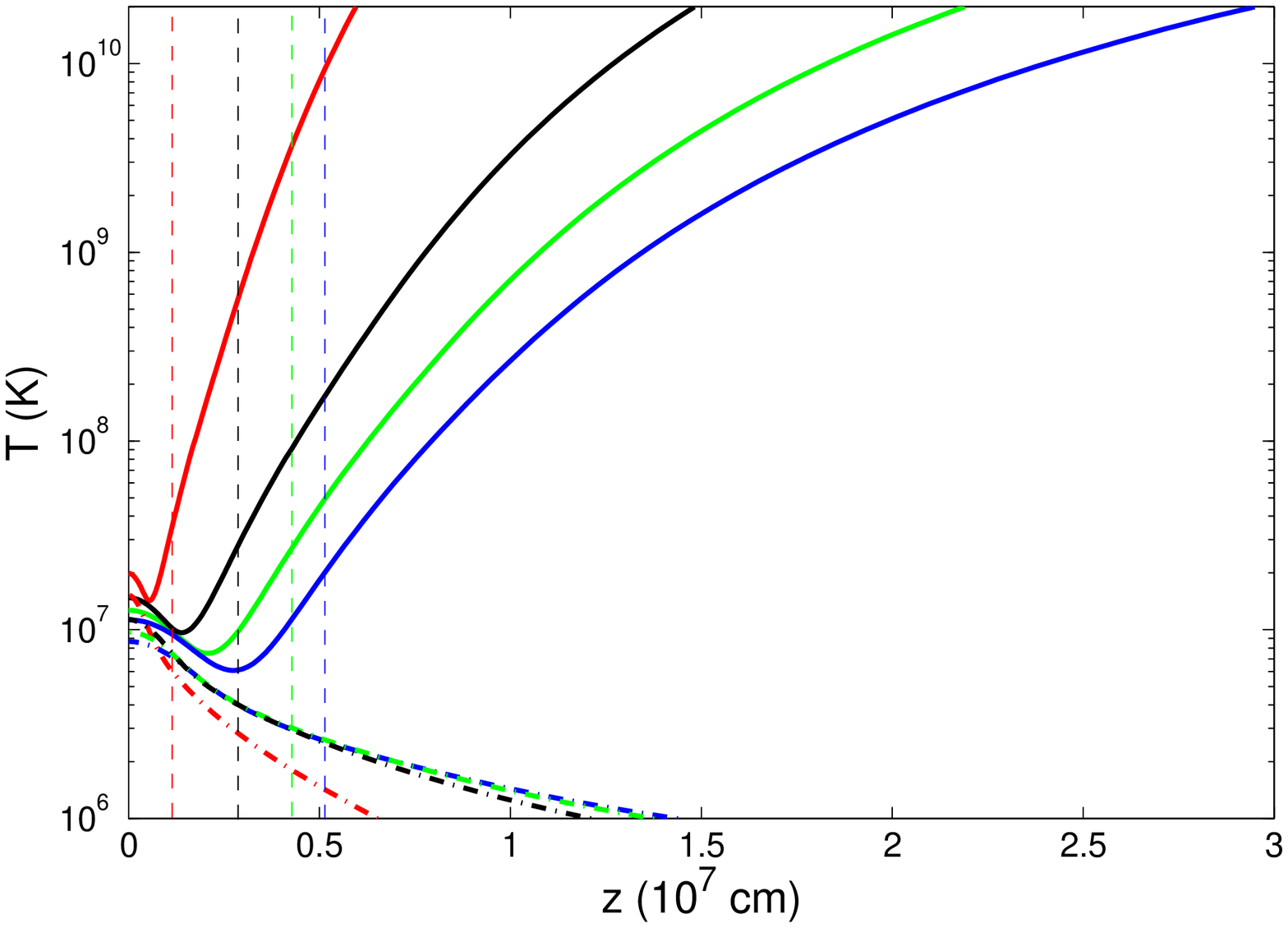}
\caption{Annuli spectrum and vertical structure due to the single power law $\zeta=0.03$ dissipation profile from equation (\ref{dis2}). The top panel shows emergent spectrum from annulus at $r/r_{\rm g}=40$ (blue), $30$ (green), $20.2$ (black) and $9.67$ (red). The bottom panel plots the gas (solid) and radiation (dot-dash) temperatures of the models, along with the locations of the corresponding scattering photospheres (dash). Note that \cite{dav05} obtained qualitatively similar high energy spectrum by manually putting $50$ percent of the dissipation outside of the scattering photosphere.}
\label{fig:z0.97}
\end{figure}

Replacing the dissipation profile with the single power law forms given by equation (\ref{dis2}) results in vastly different annuli spectra for the same $L/L_{\rm Edd}=0.8$ disk. In this case, we have by construction put more dissipation per unit mass into the low surface density upper layers. The annuli spectra now display clear high energy tails, as shown in Figures \ref{fig:z0.9} and \ref{fig:z0.97}.

\begin{figure}
\includegraphics[width=9cm]{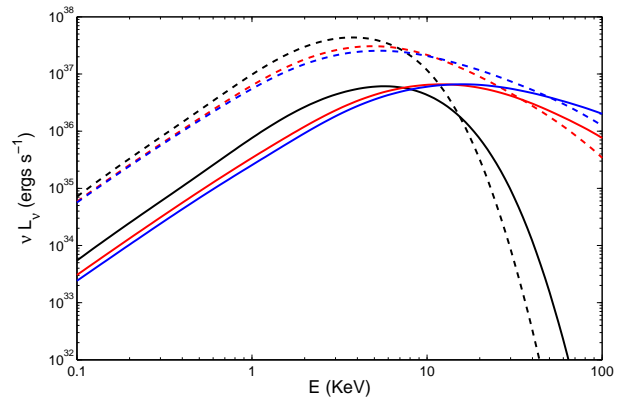}
\caption{Disk integrated spectra for the broken power law (black) and single power law (red, $\zeta=0.1$; blue, $\zeta=0.03$) dissipation prescriptions from equations (\ref{dis1}) and (\ref{dis2}), respectively. The dashed and solid curves respectively represent a disk observed face on and nearly edge on. In both cases, we have a $L/L_{\rm Edd}=0.8$ disk around a $6.62$ solar mass black hole.}
\label{fig:diskspec}
\end{figure}

Finally, Figure \ref{fig:diskspec} shows total disk integrated spectra using each of the three dissipation prescriptions while holding the disk parameters the same. The single power law dissipation profiles (dashed) put more energy per unit mass into the upper layers, resulting in spectra with strong high energy tails.

\subsection{Dissipation and Photospheres}
\begin{figure}
\includegraphics[width=9cm]{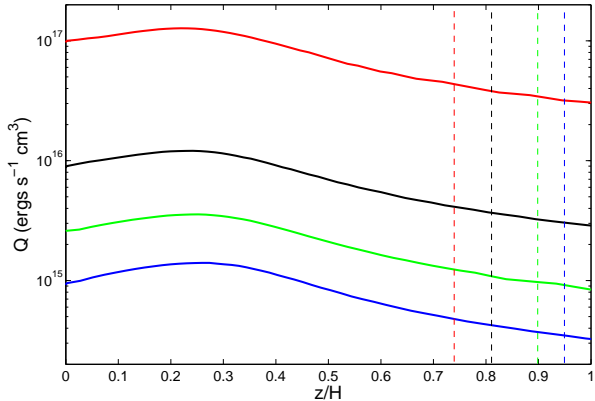}
\caption{Representative local dissipation rate as a function of height (normalized to disk scale height) for selected annuli models using the single power law $\zeta=0.1$ dissipation prescription. The solid curves correspond to $r/r_{\rm g}=40$ (blue), $30$ (green), $20.2$ (black) and $9.67$ (red). The vertical dashed lines once again indicate the locations of the scattering photospheres.}
\label{fig:z0.9q}
\end{figure}

The numerical results confirm our theoretical estimate regarding the photospheres. Figure \ref{fig:z0.9q} illustrates that the dissipation profile peaks do indeed remain at roughly the same $z/H$ (although $H$ is modified from equation (\ref{hr}) and varies with $r$ since our code took relativistic effects into account). Moreover, the scattering photospheres move toward regions of higher dissipation as we decrease $r$, so that the fractional accretion power dissipated around the photosphere (and in the outer layers) increases as we approach the black hole. 

Before moving on, we stress that the functional form of the dissipation profile deep inside the disk is sensitive to details of the thermal and magnetic pressures, but does not actually affect the photon output as the spectral shape really only depends on the dissipation per unit mass near the photospheres. In other words, the fact that the dissipation profiles peak slightly above instead of at the mid-plane is in some sense a detail that is mostly irrelevant to the spectra.

\section{Physics of Annuli Spectra Evolution}
We saw in the previous section that the two classes of dissipation profiles we used resulted in vastly different spectra. The more artificial unbroken power law dissipation profiles represented by equation (\ref{dis2}), which by construction put more energy per unit mass into the low surface density regions, gave rise to extended high energy spectral tails. On the other hand, the broken power law prescription from equation (\ref{dis1}) failed to produce a high energy spectral tail even at the remarkably low midplane surface density of $801 \rm g/cm^2$, but instead evolved from a modified black body shape directly into spectra characteristic of saturated Compton scattering as we decreased $\so$ while increasing $\O$ and $\teff$. 

Comparing Figure \ref{fig:z0.8} with either Figure \ref{fig:z0.9} or \ref{fig:z0.97}, the temperature structures resulting from the two sets of profiles indicate two important and related differences. First, in the region right below the scattering photosphere where the atmosphere is marginally optically thick to electron scattering, the single power law dissipation prescriptions result in sharper upward gas temperature gradients compared to the broken power law version, largely because the former puts more power into the upper layers. Second, the same extra power also drives the maximum gas temperature for models resulting from the unbroken power law dissipation profiles much higher. 

To make our understanding more quantitative, we compute an approximate Compton-$y$ parameter and associated photon power law index. We do not include relativistic effects in this simple estimate. In an atmosphere with vertical temperature stratification, we define a $y$-parameter in terms of the scattering optical depth $\te$. If the atmosphere is optically thin to scattering, we have
\B
y=\int_0^{\tau_{\rm *}}\frac{4\kb\tg d\te}{\me c^2}=\int_{z_{\rm max}}^{z_{\rm *}}\frac{4\kb\tg\ks\rho dz}{\me c^2},
\label{ythin}
\E
where $\tau_{\rm *}$ is the scattering optical depth where the up-scattering process effectively begins, and $z_{\rm *}$ is the corresponding physical height. We integrate inwards from the disk surface so that $z_{\rm max}$ is the maximum height of the annulus. For an optically thick medium, $y$ is the sum of equation (\ref{ythin}) and a $\te>1$ part:
\B
y=\int_0^1\frac{4\kb\tg d\te}{\me c^2}+\int_1^{\tau_{\rm *}}\frac{4\kb\tg 2\te d\te}{\me c^2}.
\label{ythick}
\E
The above expressions give the integrated or effective $y$-parameter that an observer sees by looking down a depth $\tau_{\rm *}$ into the disk. Within the Kompaneets framework, Comptonization of cool source photons by a hotter electron gas should result in an intensity spectrum of the form
\B
I_{\rm\nu}\propto\nu^{m+3},
\E
where 
\B
m+3=\frac{3}{2}\left(1\pm\sqrt{1+\frac{16}{9y}}\right)
\E
is a real constant and we have written the power-law index of $I_{\rm\nu}$ in the standard notation. For the up-scattering process under consideration, the $y$-parameter is usually significantly less than $1$ and the minus sign is generally appropriate. 

For a given converged annulus model with a prominent high energy spectral tail, we integrate equation (\ref{ythick}) from the disk surface downward and compute $m+3$ as a function of height. We then identify the height $z$ at which the power-law index approximately matches the measured slope of the high energy tail produced by the TLUSTY spectral calculation. This then gives us the integrated (or effective) $y$-parameter of the atmospheric segment that produced the power-law like spectrum along with the an idea of the depth at which the up-scattering process that led to the high energy tail began. 

We use the $r/r_{\rm g}=20.2$ annulus from the $\zeta=0.03$ run to briefly illustrate the above process. For this model, we find that $m+3$ is approximately equal to the estimated high energy tail slope of $-2$ at $\tau_{\rm *}\approx 3$ with an integrated $y$-parameter of about $0.39$. This means that the Compton up-scattering region has to extend down to about $\tau_{\rm *}\approx 3$ in order to produce the model's high energy tail, which is a very sensible result since this is approximately the depth where the upward gas temperature gradient is the steepest. Note that the photon spectrum does not display an exponential cut-off within the frequency range of our calculations but instead has a slightly concave down shape due to the vertical temperature stratification. In an isothermal medium, the photons would eventually reach the gas temperature and the up-scattering would stop. However in our model atmospheres the photons continuously move upward into regions of even higher temperature. The atmosphere becomes increasingly optically thin as photons propagate towards the disk surface and the up-scattering efficiency decreases, resulting in the continuously decreasing (or increasingly negative) spectral slope with photon energy.

In the case with the single power law dissipation profiles, the spectral forming region around the scattering photosphere is able to significantly Compton up-scatter photons from the peak of the lower energy blackbody spectrum produced deeper within the disk (below the gas temperature minimum where electrons and photons decouple). On the other hand, the broken power profile does not heat the photospheric region sufficiently and hence most lower energy photons red-ward of the thermal spectra maximum are up-scattered into the blackbody peak. This then results in the saturated Compton scattering spectrum from the models based on equation (\ref{dis1}) as the height with the lowest gas temperature (where the up-scattering begins) is still very optically thick to scattering.

It is worth noting that the region between the gas temperature minimum and the scattering photosphere is largely optically thick to electron scattering for models based on any of our dissipation prescriptions, which at first glance might lead us to expect saturated Compton scattering spectra in all cases. To understand why this does not occur, we stress that the physical reasoning in the previous paragraph is only applicable to optically thick scattering dominated photosphere having an outward rising electron temperature gradient. Here it is the marginally optically thick region right below the scattering atmosphere that controls the spectral shapes and, given the right conditions, a significant high energy tail will result even if the entire scattering dominated region (between the electron temperature minimum and the scattering photosphere) is highly optically thick. In contrast, an unstratified isothermal medium would give rise to a powerful high energy tail only if the entire scattering dominated region is not so optically thick that the photons can thermalize with the electrons before exiting. In this sense, it is really the degree of deviation of the electron temperature structure in the scattering dominated region from that of an isothermal atmosphere that determines whether the spectrum will develop a powerful tail when the medium is overall very optically thick to scattering.

\section{Discussion}
We have constructed, via calculations that self-consistently incorporate disk vertical structure, radiative transfer and physically motivated dissipation prescriptions, annuli models that have an inner region between where the gas temperature decreases outwards plus an outer layer where the temperature increases, potentially sharply, with height. Such a vertical structure can be viewed as a hot corona covering an optically thick cold disk. Variations upon this basic geometry are commonly invoked to explain and fit the high energy emissions from black hole X-ray binaries. Our results indicate that the corona, if it exists in nature, may not be isothermal and that it is the sharpness of the temperature gradient in this electron scattering dominated region that determines the emergent photon spectrum.

The locations of the scattering photosphere and gas temperature minimum move vertically inward as we decrease $r$ as predicted by our estimate in Section 3. In particular, this means that our self-consistent calculation may produce a geometry where, due to strong Comptonization, the disk becomes truncated as the radius decreases and is completely replaced by the coronal structure. This has been considered as another potential geometry underlying the observed very high state spectra \citep{dk06}. The difference between this scenario and one where the disk extends down to the last stable orbit (presumably the models shown in Figures \ref{fig:z0.8}, \ref{fig:z0.9} and \ref{fig:z0.97}) may represent the distinction between the so-called extreme very state and very high state geometries as suggested by, for example, \cite{kd04}.

We used dissipation prescriptions that are at least qualitatively consistent with the \cite{hir08} simulations to produce powerful high energy spectral tails. These dissipation profiles put more energy per unit mass into the low surface density upper layers to give rise to sharply rising gas temperature gradients. The more aggressive $\zeta=0.03$ case actually resulted in a full disk spectrum which is a good match for the steep power law observations. For example, the steep power law state observed in well-studied black hole X-ray binary systems such as XTE J1748-288 ($\Gamma\approx 2.92$) \citep{rtb00, mr06} and XTE J1550-564 ($\Gamma\approx 2.82$)\citep{dk06, mr06} have spectral index $\Gamma-2$ close to but less than $1$, where $\Gamma$ is defined such that $\nu F_{\rm\nu}\propto\nu^{-(\Gamma-2)}$.  This is in approximate quantitative agreement with the dashed blue curve (face on disk using $\zeta=0.03$) in Figure \ref{fig:diskspec}. However, our edge-on full disk spectra for both single power law dissipation prescriptions peak at an energy too high compared to observations. Nonetheless, our results therefore indicate that dissipation physics may explain the observed $\Gamma$ in the X-ray bands of some SPL observations.

We should point out a numerical caveat in our work. We had to manually impose a cut-off to the single power law dissipation profile at $\Sigma<10^{-6}\so$ in order to obtain energy conservation. Without the cut-off, our calculations would only capture about $80\%$ to $85\%$ of the expected radiative flux (given by $\sigma\teff^4$) because the computational domain is limited by convergence issues to a minimum surface density floor of $10^{-3} \rm g/cm^2$. However, the atmosphere becomes so tenuous at such low surface densities that the material there perhaps should no longer be considered part of the accretion disk. The gas temperature at the surface density minimum (which is also the maximum $\tg$ reached by the calculation) is so high that the particles beyond that height likely have enough thermal energy to escape the black hole, especially in high Eddington ratio systems. This indicates that the missing $15\%$ to $20\%$ of the expected radiative flux, which corresponds to power dissipated at heights with surface density lower than $10^{-3} \rm \ g/cm^2$, may actually go into something other than photons, mostly likely a thermally driven wind. A calculation with an even lower dissipation power law index $\zeta$ would capture even less of the expected flux within the computational domain. It is therefore possible that further increasing the dissipation per unit mass in the disk upper layers will simply result in dissipating a larger fraction of the accretion power into a wind instead of a more prominent high energy spectral component. This scenario may physically limit the ability of models like ours, where the dissipation of turbulent energy simply heats a thermal electron population, to self-consistently produce a spectrum with an X-ray tail that is significantly harder (lower $\Gamma$) than what we have so far. This means that alternative physics input such as a non-thermal electron energy distribution may be necessary to reproduce some SPL observations where $\Gamma-2$ closer to $0.5$. This case is particularly plausible given that the spectral tails can extend out to several hundred keV or more in the soft, hard \citep{lw05} and very high states. 

\section{Conclusions}
We have self-consistently computed the vertical structure and emergent spectra of accretion disk annuli models spanning a range of surface densities, effective temperatures and orbital frequencies. We focused on the effects of three different local dissipation prescriptions. We also incorporated simulation-based magnetic acceleration support against gravity.

The first dissipation profile is a power law as a function of fractional surface density, with a break at $\Sigma/\so=0.11$. This form fits the horizontal and time averaged dissipation profile from the local shearing box simulations of \cite{hir08}. The resulting annuli spectrum transitions directly from that of a modified blackbody to one characteristic of saturated Compton scattering as we decrease $\so$ while increasing $\teff$ and $\O$. The full spectrum ($L/L_{\rm Edd}=0.8$) thus constructed is largely thermal and shows a hint of the saturated Comptonization feature. The second and third dissipation profiles are unbroken power laws with artificially chosen power law indexes but are still in qualitative agreement with the simulation results when viewed as a function of height. These prescriptions put more power per unit mass into the disk upper layers compared to the broken power law case and resulted in annuli spectra with significant high energy tails, which are prominent even in the full disk spectra. In particular, the more aggressive $\zeta=0.03$ models gave rise to a disk spectrum that is in approximate quantitative agreement with some SPL observations.

Our annuli vertical structure results are consistent with theoretical expectations. As a function of height, the positions of the dissipation profile peaks were indeed tied to the disk scale height, remaining at roughly constant $z/H$ as we varied annuli parameters. Moreover, the photospheres moved inward as a function of $z/H$ as we increased $\teff$ and $\O$ while decreasing $\so$, bringing more dissipation to the spectral forming region. With both dissipation profiles, we found that the disk thickness decreased as we approached the black hole, raising the possibly of a truncated inner accretion flow.

While our results are promising, we also saw evidence that increasing dissipation in the disk upper layers may not necessarily lead to a more powerful high energy spectral component. Instead, a larger fraction of the accretion power would go into driving a thermal wind. In addition, non-thermal electron distributions, which we did not consider, are probably necessary to truly understand the full range of high energy spectral shapes observed in black hole X-ray binary systems.


We acknowledge support from the National Science Foundation under grant
AST-0707624. The authors greatly appreciate the generous help from and insightful discussions with S. Davis and I. Hubeny. We are also grateful to C. Done for illuminating conversations regarding both observational and theoretical aspects of accretion disk studies. In addition, we thank N. Shabaltas, M. Block and M. Van Adelsberg for their assistance on numerical techniques. Finally, we learned a great deal from J. Ling about gamma ray observations of black hole X-ray binaries.

\end{document}